\documentclass[%
 reprint,
 floatfix,
superscriptaddress,
 amsmath,amssymb,
 aps,
pre,
]{revtex4-2}

\usepackage{epsfig,dsfont,amssymb,amsmath,amsthm,amsfonts,amsbsy,mathrsfs}
\usepackage{graphicx}
\usepackage{color}
\usepackage{bm}
\usepackage{multirow}
\usepackage{natbib}

\usepackage{mathbbol}

\newcommand{\BEQ}{\begin{equation}}
\newcommand{\EEQ}{\end{equation}}
\newcommand{\BEA}{\begin{eqnarray}}
\newcommand{\EEA}{\end{eqnarray}}

\usepackage{tikz}
\usetikzlibrary{arrows,shapes,chains}

\begin{document}
\preprint{APS/123-QED}

\title{The most probable path of Active Ornstein-Uhlenbeck particles}

\author{Andrea Crisanti}
\affiliation{Dipartimento di Fisica, Sapienza Universit\`a di Roma
Piazzale  A.  Moro  2,  I-00185  Rome,  Italy.}

\author{Matteo Paoluzzi}
\email{matteopaoluzzi@ub.edu}
\affiliation{Departament de Física de la Mat\`eria Condensada, Universitat de Barcelona, C. Martí Franqu\`es 1, 08028 Barcelona, Spain.}

\date{\today}

\begin{abstract}
Using the path integral representation of the non-equilibrium dynamics, we compute the most probable path between arbitrary starting and final points, followed by an active particle driven by persistent noise. We focus our attention on the case of active particles immersed in harmonic potentials, where the trajectory can be computed analytically.
Once we consider the extended Markovian dynamics where the self-propulsive drive evolves according to 
an Ornstein-Uhlenbeck process, we can compute the trajectory analytically with arbitrary conditions on position and self-propulsion velocity. We test the analytical predictions against numerical simulations and we compare the analytical results with those obtained within approximated equilibrium-like dynamics.
\end{abstract}

\maketitle

\section{Introduction}




As a general feature, active particles perform persistent random walks whose characteristics, i.e., typically the persistence length or persistence time, are specific to the system of interest \cite{marchetti2016minimal,Marchetti13,PhysRevX.12.010501,Bechinger17}. 
In the case of the so-called Active Ornstein-Uhlenbeck particles (AOUp) \cite{maggi2015multidimensional,Fodor16,Szamel14,Farange15,Fily12,PhysRevE.98.020604}, the persistent random walk results from the action of a noise that is exponentially correlated in time. 
The stochastic dynamics of dynamical systems driven out-of-equilibrium by exponentially correlated noise has been largely investigated in the past decades \cite{Hanggi95}.
Their study has been performed by considering different techniques ranging from the Fokker-Planck equation \cite{Hanggi95,fox1983correlation} to the path integral representation of the dynamics \cite{PhysRevA.41.644,PhysRevA.41.657,PhysRevA.42.1982}, and from a single degree of freedom up to field theoretical models \cite{sancho1998non,maggi2022critical,PhysRevE.105.044139}.
Exponentially correlated noise reproduces quite well different situations in active matter, as in the case of the dynamics of passive beads in active baths \cite{PhysRevLett.84.3017,Maggi14,maggi2017memory}, the transport properties of passive objects in numerical simulations \cite{angelani2010geometrically,paoluzzi2016shape}, and the critical dynamics of scalar active systems \cite{maggi2022critical}. At the theoretical level, early studies showed that the dynamics of Active Brownian particles can be recast into non-Markovian dynamics characterized by a persistent noise \cite{Fily12}. 

Although some features of active systems can be rationalized in terms of effective equilibrium pictures \cite{Fodor16,Farange15,PhysRevResearch.2.023207,PhysRevE.103.032607}, in many situations that are usually those where Active Matter develops novel physics, it is not possible to ignore the non-equilibrium nature of the microscopic dynamics \cite{o2022time,PhysRevLett.123.238003,dal2019linear}. For instance, if we are interested in the escaping strategies performed by active particles for climbing the local minima of external potentials, differently from particles in contact with a thermal bath, the details of the potential and not only the height of the
energy barriers matter \cite{PhysRevLett.122.258001}. This is because the self-propulsion fixes the scale of the maximum force that the particle can exert for climbing the potential barrier and eventually to escape from a local minimum (in AOUp this is related to the inflection point of the potential \cite{PhysRevA.41.657,caprini2019active}). This fact has also important consequences in many situations, as in the case of the glass transition \cite{paoluzzi2022motility} or when active particles move in heterogeneous media \cite{PhysRevLett.124.118002}. 
These are typical situations where we can not ignore the non-equilibrium features of the active motion so one has to develop other techniques for describing stationary properties of the active system \cite{Woillez_2020}.

Among other approaches for addressing equilibrium and non-equilibrium dynamics, 
the path integral representation of the stochastic dynamics remains a powerful tool. This is because, when the strength of the noise is small, the path integral is dominated by the trajectory that extremizes the corresponding dynamical action, i.e., we can perform the saddle-point approximation. The trajectories obtained in this limit are usually called
instantons \cite{caroli1981diffusion,brezin1977perturbation,lopatin1999instantons}.
In performing instantonic computations, the natural set-up is to prepare the system in un minimum in the infinite past and compute the escape rate to another minimum that is another stationary point in the infinite future. On the other hand, once we recast the dynamics in terms of path integrals, one can formally compute the probability of trajectories connecting arbitrary points on a finite time scale. In this way, the saddle-point approximation of the path integral returns the most probable trajectory followed by the particles for moving between arbitrary initial and final points. 

Looking at the case of exponentially correlated noise, the computation of instantons has been performed intensively in the past  \cite{bray1989instanton,PhysRevA.41.644,PhysRevA.41.657,PhysRevA.42.1982}. However, 
analytical progress can be done only n the small and large $\tau$ limit \cite{bray1989instanton}, with $\tau$ indicating the noise correlation time. Moreover,  most of the attention has been devoted to the computation of the escape rate rather than the properties of the most probable trajectories on a finite time interval.

In the case of Active Matter, the knowledge of the morphology of the most probable trajectories followed by active particles might elucidate further not only the mechanisms of active escaping but also other general features of active agents, as possible optimal strategies in moving in complex environments. The computation of the most probable path in the case of non-interacting Active Brownian particles has been studied only recently \cite{yasuda2022most}. In the case of activated escape from confining potentials, it is possible to compute optimal escape paths between stationary configurations in the small noise limit \cite{PhysRevLett.122.258001}. However, the computation of the most probable path with arbitrary initial and final conditions in absence of thermal noise remains poorly studied.

In the present work, we are going to address the general question about the most probable path between two arbitrary points in the case of non-Markovian dynamics that is correlated on a finite timescale.
We thus specialize our computation to the case of AOUPs where the noise is exponentially correlated on a single time scale. As a concrete example that is analytically tractable, we consider 
the active particle confined by harmonic potentials.
From the analytical computation of the average trajectories followed by the system, and using the persistence time $\tau$ as a control parameter, we document how the morphology of the typical trajectory changes from Brownian-like to quasi-ballistic as $\tau$ increases and we test the analytical predictions against numerical simulations. We also compute the extreme path of the extended dynamics, where we can take into account also the boundary conditions on the self-propulsion force. In this case, we can compute the most probable trajectories conditioned by boundary conditions on the self-propulsive force. In this way, we can make in relation the typical shape of the trajectory with the value of the self-propulsion at the endpoints. 

Finally, we compare the trajectories of the actual dynamics with those obtained within effective equilibrium approaches. Although the stationary distribution in the effective equilibrium picture matches the analytical distribution, we show that they differ considerably almost everywhere in the finite time interval where the dynamics takes place. We recover a quantitative agreement between the two dynamics only on small times when $\tau \to 0$.

\begin{figure}[!t]
\centering\includegraphics[width=.45\textwidth]{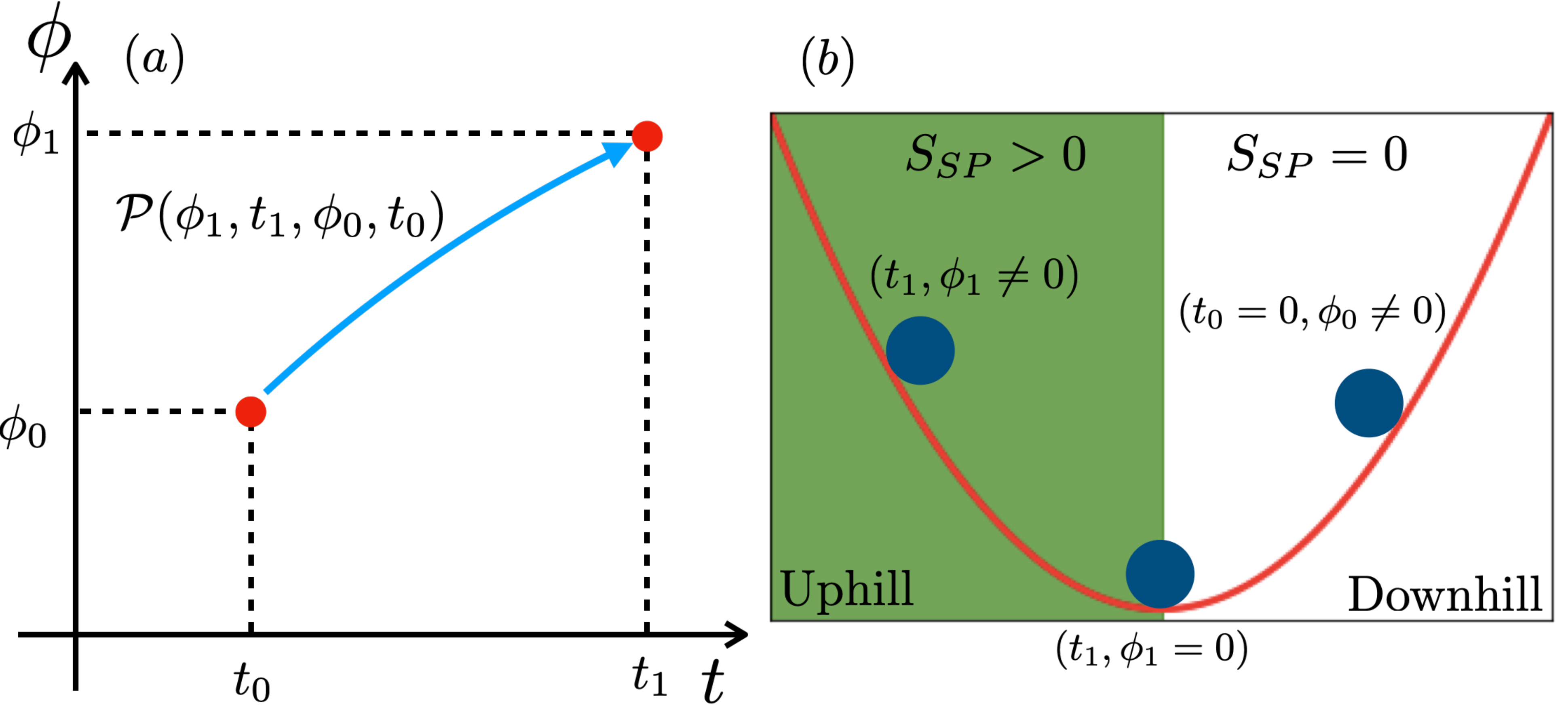}
\caption{(a) 
We consider a particle located at the initial time $t_0$ in $(t_0,\phi_0)$
that reaches the final point $(t_1,\phi_1)$ at the time $t_1$. 
Once we introduce the probability $\mathcal{P}(\phi_1,t_1,\phi_0,t_0)$ as a sum over the paths connecting the two points, the most probable path corresponds to the saddle-point solution of the path integral. (b) The most probable trajectories return the value $S_{SP}$ of the dynamical action at the saddle-point. For downhill trajectories, i.e., starting from $\phi_0\neq0$ and ending in $\phi_1=0$, the dynamical action vanishes so that $S_{SP}=0$. In the case of the uphill trajectories connecting $\phi_0=0$ with $\phi_1\neq0$, $S_{SP}$ returns a finite positive value.}
\label{fig:fig1}
\end{figure}

\section{Dynamics}
We indicate with $\phi$ the degree of freedom of the system, i.e., in our case particle's position, the time evolution of $\phi$ results
from the competition between a deterministic force $f(\phi)$ and a stochastic force that we indicate with $\varphi$. The equation of motion is
\begin{align} \label{eq:dyn1}
    \dot{\phi} = f(\phi) + \varphi \; .
\end{align}
in this picture, without loss of generality, we set the mobility of the particles $\mu\!=\!1$ (we are working in the low Reynolds numbers regime where inertia is negligible). 
We assume the deterministic force generated by a conservative field, i.e., $f(\phi) = -V^\prime(\phi)$ (dots and primes indicate time derivative and derivative with respect to $\phi$, respectively). 
In this way, any non-equilibrium contribution is due to the fluctuating force $\varphi$ representing the action of fast degrees of freedom on the relevant variable $\phi$ we are interested in. In the following, we assume that  $\varphi$ is a Gaussian noise with zero mean so that the statistical properties of the fluctuating force are specified by the second cumulant while higher-order cumulants are zero. 
In general, one has
\begin{align} \label{eq:memory}
    \langle \varphi(t) \varphi(t^\prime) \rangle = 2 D \,  K(|t-t^\prime|/\tau) \; ,
\end{align}
with the parameter $D$ measuring the strength of the noise and $\tau$ the persistence time. As discussed in Ref. \cite{PhysRevA.41.644},  $K$ can be any arbitrary normalizable function. When we specialize the model, we will consider the simplest case of the exponentially correlated noise since it is relevant for the study of active systems
\begin{align} \label{eq:kappa}
K(t) = \frac{1}{\tau} e^{-|t|/\tau}  \; ,
\end{align}
in this case, the dynamics of the fluctuating force can be represented using the following Ornstein-Uhlenbeck process
\begin{align} \label{eq:oup}
    \dot{\varphi} = -\tau^{-1} \varphi + \zeta 
\end{align}
with the noise $\zeta$ Gaussian with zero mean and noise strength $2 D/\tau^2$.
In the next section, to introduce the formalism, we will start working with the generic non-Markov process expressed by Eq. (\ref{eq:dyn1}) with the fluctuating force $\varphi$ whose statistical properties are given by Eq. (\ref{eq:memory}).

\section{Path Integral Formalism}
We want to compute the probability $\mathcal{P}(\phi_1,t_1,\phi_0,t_0)$ 
of a trajectory connecting the starting point $\phi_0\equiv \phi(t_0)=(t_0,\phi_0)$ 
with the ending point $\phi_1\equiv \phi(t_1)=(t_1,\phi_1)$, as sketched in Fig. (\ref{fig:fig1}a). 

We now introduce the path integral formalism that is suitable for computing trajectories connecting arbitrary points in the case of a generic potential $V$. We anticipate that
we will specilize our computations in the next sections on the case of a particle in the harmonic trap, as illustrated in Fig. (\ref{fig:fig1}b).  
This is because, in the case of harmonic potentials, the dynamical action is quadratic so that the saddle-point approximation provides the analytical solution of the path integral.

If the potential $V$ contains arbitrary non-linear interactions, 
the saddle-point approximation holds only in the small noise limit, i.e., $D\to 0$. In Active Matter, since we can perform a mapping at the single-particle level between the strength of the noise and self-propulsion through  $D=v^2 \tau$ with $v$ the self-propulsion velocity \cite{Fily12}, the small noise limit implies a small self-propulsion velocity limit. However, we stress that for quadratic potential there are no restrictions on $D$ and thus on $v$.

We start with considering the non-equilibrium dynamics specified by Eq. (\ref{eq:dyn1}) within a finite time interval $t \in [t_0,t_1]$. We first notice that the equation provides a one-to-one map $\phi(t) \to \varphi(t) $ so that, once we specify the initial condition $\phi(t_0)=\phi_0$, the solution for a given $\varphi$ is unique. Now, we want to fix also the final condition $\phi(t_1)=\phi_1$ and compute the most probable path connecting these two points. Usually, this program can be done using the corresponding Onsager-Machlup action. In the following, we will employ a different program based on response fields that provide additional degrees of freedom for fixing the initial and final conditions. We will see how the condition on the endpoint naturally emerges through the initial condition on the auxiliary field.
\begin{figure}[!th]
\centering\includegraphics[width=.4\textwidth]{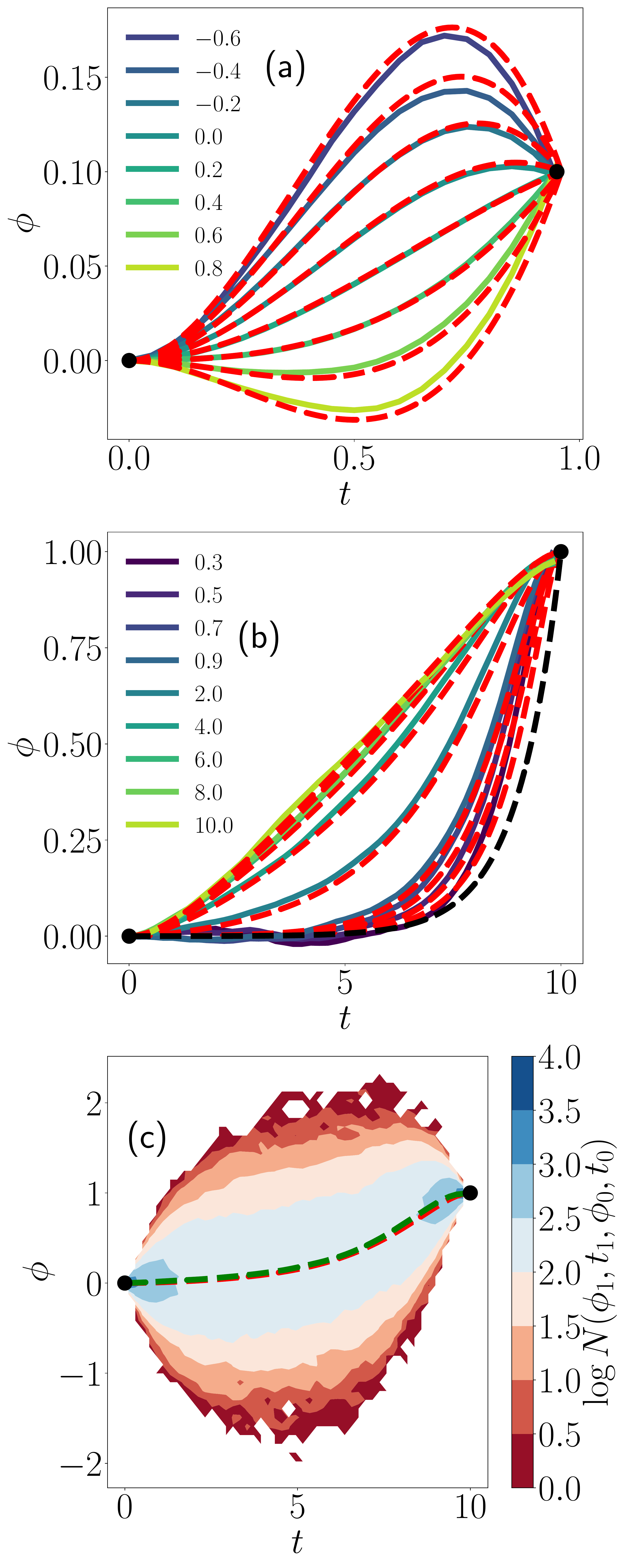}
\caption{Active Ornstein-Uhlenbeck particle in harmonic potential. (a) The most probable path connecting $(t_0=0,\phi_0=0)$ with $(t_1=0.95,\phi_1=0.1)$ ($\tau=2$) for different values of $\varphi_1 \in [-0.6,0.8]$ (see legend). Colors from violet to yellow refer to numerical simulations, dashed red curves are the analytical solution. (b) 
The most probable path from $(0,0)$ to $(10,1)$ for different values of $\tau$ (see legend). Red dashed curves are the theoretical prediction Eq. (\ref{eq:solution}). The dashed black curve is the Brownian limit Eq. (\ref{eq:br}).
(c) The color map indicates the number of trajectories connecting $\phi_0$ with $\phi_1$ (from the numerical data, $\tau=2$)
The dashed green curve is the average path, the dashed red curve is the theoretical prediction.}
\label{fig:fig2}
\end{figure}

We set the framework by introducing the path probability in the following way
\begin{align}
    \mathcal{P}(\phi_1,t_1,\phi_0,t_0) = \int_{\phi_0}^{\phi_1} \mathcal{D}[\phi] \mathcal{D}[\varphi] P[\varphi] \delta\left( \phi - \phi_\varphi \right) \; .
\end{align}
Where the path integral is extended to all the trajectories connecting $\phi_0$ with $\phi_1$ and $t\in[t_0,t_1]$. We have indicated with $\phi_\varphi$ a solution of Eq. (\ref{eq:dyn1}). 
We can perform the standard change of variable $\phi_\varphi \to \phi$ that brings to
\begin{align}
    \delta\left( \phi - \phi_\varphi \right) &= \left|\det \mathcal{J} \right| \delta\left( \dot{\phi} -f(\phi) -\varphi\right) \\ \nonumber 
    \mathcal{J} &\equiv \frac{\delta \eta }{\delta \phi} 
    \; .
\end{align}
The computation of the determinant $\det \mathcal{J}$ can be found in Refs. \cite{PhysRevA.41.644,gozzi1983functional}. 
For our purpose, since $\det \mathcal{J}$ does not depend on the strength of the noise $D$, we can neglect its contribution. 
For representing the delta-functional, we introduce the response field $\hat{\phi}$. Once we do that, we can average over $\varphi$ so that the path probability becomes 
\begin{align} \label{eq:path_MSR}
    \mathcal{P}(\phi_1,t_1,\phi_0,t_0) = \int_{\phi_0}^{\phi_{1}} \mathcal{D}[\phi] \mathcal{D}[\hat{\phi}] \, e^{-S[\phi,\hat{\phi}]}
\end{align}
where we have introduced the dynamical action $S[\phi,\hat{\phi}]$ that is
\begin{align} \label{eq:dyn_action} 
    S[\phi,\hat{\phi}] &= \int_{t_0}^{t_1} dt dt^\prime \, \mathcal{L}[\phi,\hat{\phi}] \\ 
    \mathcal{L}[\phi,\hat{\phi}] &\equiv  - \frac{D}{2} \hat{\phi}(t) K(t,t^\prime) \hat{\phi}(t^\prime) \\ \nonumber  
    &+  \hat{\phi}(t) \left[ \dot{\phi}(t) -f(\phi) \right] \delta(t-t^\prime) \; .
\end{align}
The corresponding Onsager-Machlup action $S_{OM}[\phi]$ \cite{PhysRev.91.1505} 
can be obtained upon functional integration over response fields \cite{tauber2014critical}.
It is worth noting that the functional integration over $\hat{\phi}$ in this generic case where the noise is not delta-correlated requires the computation of the inverse operator $K^{-1}$ defined as 
\begin{align}
    \int_{t_0}^{t_1} dt^{\prime\prime} \, K^{-1}(t,t^{\prime\prime}) K(t^{\prime\prime},t^\prime) = \delta(t-t^\prime)
\end{align}
the computation in the case of exponentially correlated noise is provided in Appendix (\ref{ap:ap1}).

In the following, we are not going to compute $S_{OM}[\phi]$ but we will work with the dynamical action $S[\phi,\hat{\phi}]$.
Once we perform the replacing $\hat{\phi} \to \hat{\phi} / D$, we obtain that the path probability can be computed using the saddle-point approximation 
\begin{align} \label{eq:saddle_point} \nonumber 
    \mathcal{P}(\phi_1,t_1,\phi_0,t_0) &= \int_{\phi_0}^{\phi_1} \mathcal{D}[\phi] \mathcal{D}[\hat{\phi}] \, e^{-S[\phi,\hat{\phi}]/D} \\  
    &\simeq e^{-S[\phi_{SP},\hat{\phi}_{SP}] / D}
\end{align}
where $\phi_{SP}$ and $\hat{\phi}_{SP}$ are fixed by the self-consistency saddle-point equations
(from now on we remove the label $SP$ from the dynamical variables $\phi$ and $\hat{\phi}$ for making the notation lighter)
\begin{align} \label{eq:sp_eq}
\frac{\delta S}{\delta \phi(t)}       &= \left. \frac{1}{\hat{\phi}} \frac{d\hat{\phi}}{dt} - V^{\prime \prime} (\phi) \right|_{SP} = 0 \\ \nonumber
\frac{\delta S}{\delta \hat{\phi}(t)} &= \left. -\int_{t_0}^{t_1} dt^\prime K(t,t^\prime) \hat{\phi}(t^\prime) + \frac{d \phi}{dt} + V^\prime(\phi )\right|_{SP} = 0 
\end{align}
equations that have to be solved with the two boundary conditions $\phi(t_1)=\phi_1$ and $\phi_{0}=\phi(t_0)$. When $V$ is an arbitrary function of $\phi$, the saddle-point approximation is thus well justified by the fact that we are interested in the small noise limit (and thus large $1/D$ values). In this set-up, for instance, we can replace $V$ with a double-well and prepare the system in one of the two minima. In this case, the validity of the saddle-point approximation requires the noise strength $D$ to be small compared with the energy barrier so that we look at the rare events that produce a jump from one minimum to the other. The corresponding instantonic trajectories are those obtained by solving Eqs. (\ref{eq:sp_eq}).

We stress that the introduction of the response field 
allows us to introduce the boundary condition on the arrving point 
in a transparent way. 
To illustrate this we start with considering the dynamical action $S_{SP}$ at the saddle-point that, because of Eq. (\ref{eq:saddle_point}), takes the form
  \begin{align} \label{eq:s_sp}
     S_{SP} = \frac{D}{2} \int_{t_0}^{t_1} dt dt^\prime \, \hat{\phi}(t) K(t,t^\prime) \hat{\phi}(t^\prime) \; .
 \end{align}
For making the discussion as simpler as possible, we consider the case $t_0 \to -\infty$ and $t_1 \to \infty$ so that $\phi_0=\phi(-\infty)$ and $\phi_1=\phi(\infty)$. We immediately realize that $S_{SP}\neq0$ for non-vanishing $\hat{\phi}(s)$.
 Basically, $\hat{\phi}$ plays the role of a time-dependent external field that is identically zero for a downhill (where $\phi_1=0$) solution and different from zero for uphill trajectories (where $\phi_1 \neq 0$). 

In fact, from the solution of the saddle-point equation for $\hat{\phi}$ (see Eqs. (\ref{eq:s_sp})), it follows that $\hat{\phi}=0$ is a solution corresponding to zero noise. 
Once we plug this solution into the equation for $\phi$ we obtain the downhill motion $\dot{\phi} \!=\! -V^\prime$ and plugging $\hat{\phi}=0$ into Eq. (\ref{eq:s_sp}) we have $S_{SP}\!=\!0$. 
On the other hand, uphill trajectories can be performed only 
with the help of a non-vanishing noise that guarantees 
the drives the system towards the final state $\phi_1$ (in this case the response field acts as an external field that drives the particle uphill). This trajectory
corresponds to a non-vanishing solution for the equation of the auxiliary field $\hat{\phi}$ 
and thus $S_{SP}\!\neq\! 0$. In this case, the equation for $\phi$ is $\dot{\phi} \!=\! -V^\prime + \hat{\phi}$ with, in general, $\hat{\phi}=\hat{\phi}(\phi)$. 
As a consequence, the downhill solution will be the one with zero saddle-point action, i.e., $S_{SP}^{Down}=0$, while for the uphill solution we have $S_{SP}^{Up} > 0$, as illustrated in Fig. (\ref{fig:fig1}b) in the case of a harmonic potential.
 
 \subsection{Brownian particle in harmonic trap}
 As an example that illustrates how formalism works, let us consider the simple case of a Brownian particle in contact with a thermal bath at temperature $D$ and confined by a harmonic potential $V(\phi)\!=\!r \phi^2/2$. In this case, $K(t)\!=\!\delta(t)$, moreover,
 since the path integral is a Gaussian functional integral, 
the saddle-point approximation provides the exact solution. 
In other words, the most probable path connecting the two points coincide with the {\it classical trajectory} generated by the dynamical action, i.e., $\langle\phi(t)\rangle = \phi_{SP}(t) \equiv \phi(t)$.
 
\begin{figure}[!t]
\centering\includegraphics[width=.4\textwidth]{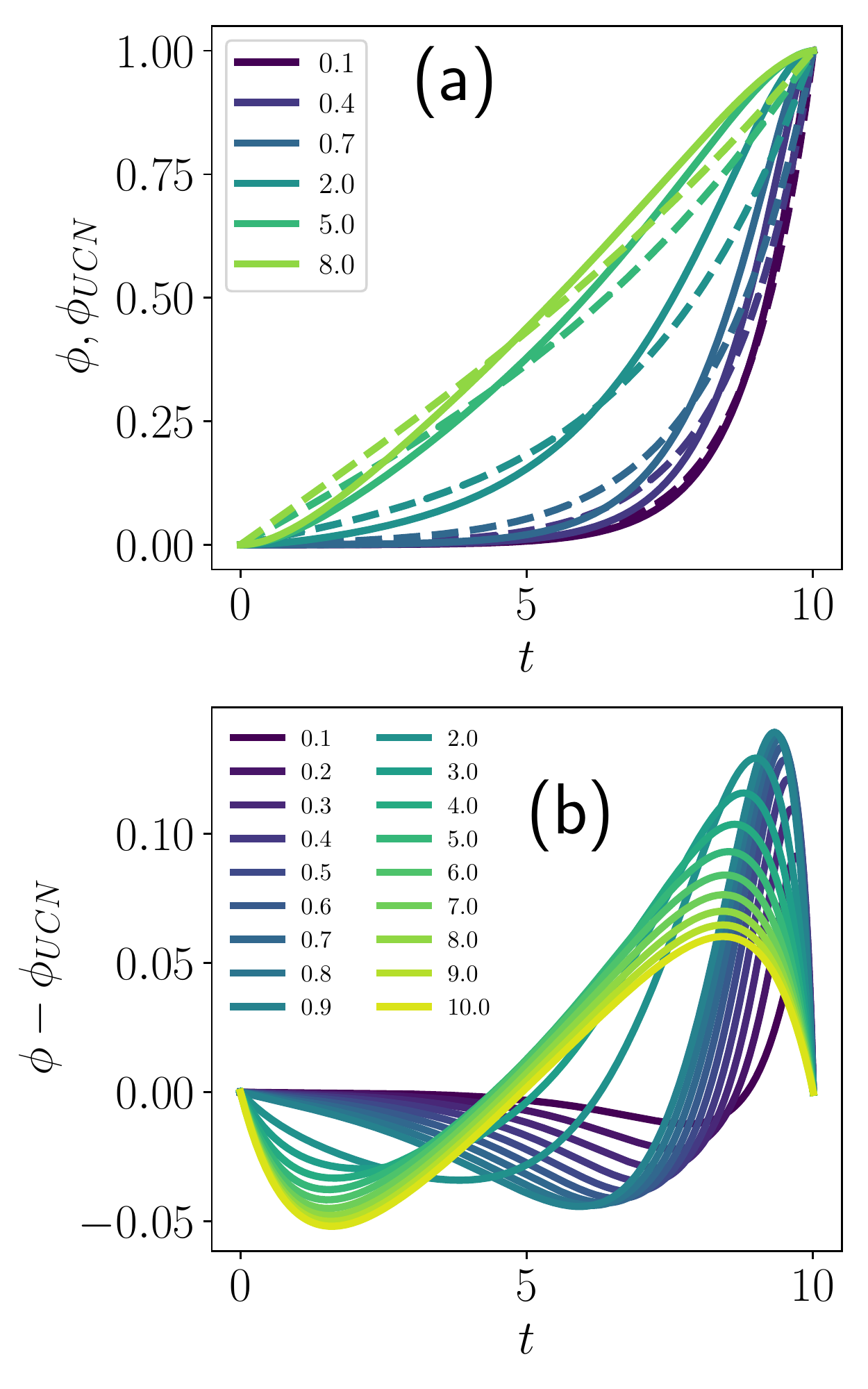}
\caption{Most probable path within the effective equilibrium action. (a) 
Solid curves are the analytical solution $\phi(t)$, dashed curves the most probable trajectories $\phi_{UCN}(t)$ within an effective equilibrium approach (increasing values of $\tau$ from violet to yellow, see legend). (b) The difference between $\phi$ and $\phi_{UCN}$ reveals the discrepancy between the two dynamics almost everywhere in a wide range of $\tau$ values.}
\label{fig:fig3}
\end{figure}

The analytical solutions of the self-consistency equations are
\begin{align}
    \hat{\phi}(t) &= \hat{\phi}_0 e^{r(t-t_0)} \\ 
    \phi(t) &= \phi_0 r^{-r(t-t_0)} - \frac{\hat{\phi}_0}{r} \sinh{\left[ r(t - t_0)\right]}
\end{align}
once we impose the condition $\phi_1 = \phi(t_1)$ we obtain the classical trajectory 
\begin{align} \label{eq:br}
    \phi(t) &= \bar{\phi}_0(t) + 
    \frac{\phi_1 - \bar{\phi}_0(t)}{\sinh{\left[ r(t_1 - t_0)\right]}} \sinh{\left[ r(t - t_0)\right]} \\
    \bar{\phi}_0(t) &\equiv \phi_0 e^{-r(t - t_0)}
\end{align}
and the action at the saddle point is
\begin{align}
    S_{SP} &= \frac{r}{4}\left( \bar{\phi}_0(t_1) - \phi_1\right)^2 (1 - \coth{ \left[ r (t_0 - t_1) \right] }) \; .
\end{align}
Fixing for convenience $t_0\!=\!0$, if we look at a downhill solution of the classical dynamics $\phi_0 \neq 0$ and $\phi_1\!=\!0$, as sketched in Fig. (\ref{fig:fig1}b), we obtain
\begin{align}
S_{SP}^{Down} = \frac{r}{4} \phi_0^2 e^{-2 r t_1} \left( 1 + \coth{ r t_1 } \right)    
\end{align}
that is non-zero at $t\!=\!0$ and zero 
in the long time limit.
The uphill solution requires $\phi_0\!=\!0$ and $\phi_1 \!\neq\! 0$ so that $S_{SP}$ is finite along this trajectory and, in the long time limit, we recover the stationary (equilibrium in this case) distribution
\begin{align}
    S_{SP}^{Up} = \frac{r}{4} \phi_1^2 \left( 1 + \coth{r t_1} \right) \xrightarrow[t_1 \to \infty]{} \frac{r}{2} \phi_1^2 \neq 0 \; .
\end{align}
so that the stationary distribution is the Boltzmann one $\mathcal{P}(\phi) \propto e^{-r\phi^2/2}$ upon replacing $\phi = \lim_{t_1 \to \infty } \phi(t_1)$.

\section{Active Ornstein-Uhlenbeck particle in harmonic trap}
We specialize our computation to the case of an active particle driven by a persistent noise and confined through a harmonic potential. We consider the one-dimensional case, however, the computation can be extended to any dimensional space.
From now on, $\phi$ represents the coordinate of the particle and $\varphi$ is the self-propulsive force. Passive objects in an active bath as colloidal beads immersed in {\it E. coli} are typical examples of AOUP particles \cite{PhysRevLett.84.3017}. Although AOUPs in the harmonic trap do not show non-equilibrium condensation phenomena as in the case of run-and-tumble and Active Brownian particles \cite{Tailleur09,solon2015active}, they still display many non-equilibrium features as manifested by a peculiar violation of the Fluctuation-Dissipation Theorem  \cite{maggi2017memory}, moreover they satisfy a modified version of the equipartition theorem \cite{Maggi14}.   
Finally, at the field theoretical level, the Gaussian theory provides a useful framework for rationalizing the behavior of critical scalar active matter \cite{maggi2022critical} suggesting that the understanding of the free theory, i.e, the counterpart of the harmonic oscillator in the field theory, might still provide non-trivial insight into the complex dynamics of Active Matter at the coarse-graining level \cite{Paoluzzi16}. Again, the harmonic case provides a useful picture for gaining insight into the early stages of active glass fluidization \cite{paoluzzi2022motility,henkes2020dense}.

In the following, we first compute the most probable path connecting $(\phi_0,\varphi_0)$ with $(\phi_1,\varphi_1)$ of the extended Markovian dynamics. Then we compute the most probable trajectory from $\phi_0$ to $\phi_1$ of the non-Markovian dynamics.

\subsection{Most probable path of the extended Markovian dynamics}
We are going to compute the most probable path that connects the initial point $\phi_0$ with self-propulsion velocity $\varphi_0$ with the endpoint characterized by $\phi_1$ and $\varphi_1$. With this aim, we start with embedding the original one-dimensional non-Markovian dynamics into a two-dimensional Markovian process so that we can write
\begin{align} \label{eq:linear_sys}
    \underline{\dot{\phi}} = -\boldsymbol{M} \underline{\phi} + \underline{\boldsymbol{\eta}} \; .
\end{align}
In our case, $\boldsymbol{M}$ is a $2\times2$ non-symmetric matrix 
(see Ref. \cite{PhysRevE.85.061127} for the case 
of two linear processes coupled to two thermostats), 
i.e., $\boldsymbol{M} \!\neq \!\boldsymbol{M}^T$, 
and $\underline{\phi}=(\phi,\varphi)$. 
However, the following results are not limited 
to the two-dimensional case, but they hold 
in arbitrary dimensions, i.e., $\underline{\phi} =(\phi_1,\phi_2,...,\phi_n)$. We thus consider 
a generic Gaussian noise vector $\underline{\boldsymbol{\eta}}$ with zero-mean
\begin{align}
    \langle \eta_{a}(t) \eta_b(t^\prime) \rangle = 2D_{ab} \delta(t-t^\prime)
\end{align}
with $a,b=1,...,n$.
In the following we adopt the notation
\begin{align}
    \underline{f}^T \boldsymbol{K} \underline{g} \equiv \sum_{a,b}  f_a(t) K_{ab}(t,t^\prime) g_b(t^\prime) \; .
\end{align}
The probability of the path that connects $\underline{\phi}_0$ with $\underline{\phi}_1$ is given by
\begin{align} \label{eq:joint}
    \mathcal{P}(\underline{\phi}_1,t_1,\underline{\phi}_0,t_0) &= \int_{\underline{\phi}_0}^{\underline{\phi}_1} \mathcal{D} [\underline{\phi}]\mathcal{D} [\underline{\hat{\phi}}] \, e^{-S[\underline{\hat{\phi}}, \underline{\phi}  ]} \\ \nonumber 
    S[\underline{\hat{\phi}}, \underline{\phi}  ] &= \int_{t_0}^{t_1} dt dt^\prime \mathcal{L}[\underline{\hat{\phi}}, \underline{\phi}] \\ \nonumber 
    \mathcal{L}[\underline{\hat{\phi}}, \underline{\phi}] &\equiv -\frac{1}{2} \underline{\hat{\phi}}^T \boldsymbol{D}\underline{\hat{\phi}} + \underline{\hat{\phi}}^T \left( \mathbb{1} \frac{d}{dt} - \boldsymbol{M} \right) \underline{\phi}  \; ,
\end{align}
with $\mathbb{1}$ indicating the $n \!\times\! n$ identity matrix.
We now specialize our computation to the case of an AOUp particle in one spatial dimension where, in the case where we consider only active driving, the matrices $\boldsymbol{M}$ and $\boldsymbol{D}$ 
take the form
\begin{align} \label{eq:noise_matrix}
    \boldsymbol{M} = 
    \begin{bmatrix}
    r & -1 \\
    0 & \frac{1}{\tau}
    \end{bmatrix} \; , \;
    \boldsymbol{D} = 
    \begin{bmatrix}
    0 & 0 \\
    0 & D 
    \end{bmatrix} \; .
\end{align}
The two-dimensional vector $\underline{\phi}$ takes the form $\underline{\phi}=(\phi,\varphi)$, we thus obtain the following dynamical action
\begin{align}
    S &= \int_{t_0}^{t_1} dt \, \mathcal{L}[\hat{\varphi},\hat{\phi},\varphi,\phi] \\ \nonumber 
    \mathcal{L} &\equiv -\frac{D}{2}  \hat{\varphi}^2  +  \hat{\phi} \left[ \frac{d}{dt} + r \right] \phi - \hat{\phi} \varphi + \hat{\varphi} \left[ \frac{d}{dt} + \tau^{-1 }\right] \varphi \; .
\end{align}

The equations of the most probable path $(\phi(t),\varphi(t))$ connecting $(\phi(t_0),\varphi(t_0))$ with $(\phi(t_1),\varphi(t_1))$ can be obtained by solving  the saddle-point equations (we set $D\!=\!1$)
\begin{align}
    \frac{\delta S}{\delta \phi (t)} &= -\frac{d \hat{\phi}}{dt}   + r \hat{\phi} = 0 \\ 
    \frac{\delta S}{\delta \varphi (t)} &= -\frac{d \hat{\varphi}}{dt}  + \tau^{-1} \hat{\varphi} - \hat{\phi}= 0 \\ 
    \frac{\delta S}{\delta \hat{\varphi} (t)} &= \frac{d \varphi}{dt}  + \tau^{-1} \varphi - \hat{\varphi}= 0 \\ 
    \frac{\delta S}{\delta \hat{\phi}(t)} &= \frac{d \phi}{dt}  + r\phi - \varphi = 0 
\end{align}
whose solutions can be written as follows (we set $t_0\!=\!0$ and, without loss of generality, $\varphi_0\!=\!\phi_0\!=\!0$)
\begin{align}
    \hat{\phi}(t) &= \hat{\phi}_0 \, e^{rt} \\ 
    \hat{\varphi}(t) &= \hat{\varphi}_0 \, e^{t/\tau} + \hat{\phi}_0 I(t)\\
    \varphi(t) &= \hat{\phi}_0 \, C(t) + \hat{\varphi}_0 \, D(t)\\
    \phi(t) &= \hat{\phi}_0 \, A(t) + \hat{\varphi}_0 \, B(t)
\end{align}
where we have defined
\begin{align}
    I(t) &\equiv  \frac{e^{t/\tau} - e^{rt}}{r - \tau^{-1}} \\ 
    A(t) &\equiv  \frac{\tau ^2 \left[ \sinh (r t)-r \tau  \sinh \left(\frac{t}{\tau }\right)\right]}{r-r^3 \tau ^2} \\
    B(t) &\equiv  \frac{\tau ^2 \left[ r \tau  \sinh \left(\frac{t}{\tau }\right)+e^{-r t}-\cosh \left(\frac{t}{\tau }\right)\right] }{r^2 \tau ^2-1} \\
    C(t) &\equiv  \frac{\tau ^2 \left[ r \tau  \sinh \left(\frac{t}{\tau }\right)-\sinh (r t)-\cosh (r t)+\cosh \left(\frac{t}{\tau }\right)\right]}{r^2 \tau ^2-1}\\ 
    D(t) &\equiv  \tau \sinh{ (\frac{t}{\tau})} \; .
\end{align}
We can write the trajectories in a compact form
\begin{align}
    \underline{\phi}(t) = \boldsymbol{G}(t) \underline{\hat{\phi}}_0
\end{align}
where we have defined the matrix
\begin{align}
\boldsymbol{G}(t) &\equiv 
       \begin{bmatrix}
    A(t) & B(t) \\
    C(t) & D(t)
    \end{bmatrix} \; .
\end{align}
As we sketched in the previous section, for fixing the initial condition on the auxiliary field $\underline{\hat{\phi}}_0$ we set the condition on the endpoint $\underline{\phi}_1 = \underline{\phi}(t_1)$
\begin{align} \label{eq:sol_2d}
    \underline{\phi}(t) = \boldsymbol{G}(t) \boldsymbol{G}^{-1}(t_1) \underline{\phi}_1 \; .
\end{align}
In Fig. (\ref{fig:fig2}a) we test the analytical solution given by Eq. (\ref{eq:sol_2d}) against numerical simulations of the dynamical process. Different curves indicate trajectories $\phi(t)$ with different boundary conditions on $\varphi$. As one can see, the analytical prediction reproduces quite well the numerical data (details about numerical simulations are provided in the Appendix (\ref{ap:ap2})).  

Here we are considering as initial condition $t_0\!=\!0, \phi_0\!=\!\varphi_0\!=\!0$.
We observe that the shape of the trajectories strongly depend on the value of the self-propulsion speed at the endpoint $\varphi_1$ at $t_1\!=\!1$ (with $\phi_1=1$).

\subsection{Most probable trajectory of the non-Markovian dynamics}
The embedding of the original dynamics into a larger space where the dynamics is Markovian allowed us to write the joint probability $\mathcal{P}(\phi_1,t_1,\phi_0,t_0,\varphi_1,t_1,\varphi_0,t_0)$ that is given by Eq. (\ref{eq:joint}). In this way, we have computed most probable paths with fixed conditions not only on position but also on the propulsive force. We are now going to marginalize the self-propulsion process $\varphi$ for obtaining the most probable path $\phi(t)$ of the non-Markov process.

We start from the stochastic differential equation for $\varphi$ given by Eq. (\ref{eq:oup}) whose solution is
\begin{align} \label{eq:sol_varphi}
    \varphi(t) &= \overline{\varphi}(t,t_0) + \eta(t) \\
    \overline{\varphi}(t,t_0) &\equiv \varphi_0 e^{-(t - t_0)/\tau }\\ 
    \eta(t) &\equiv \int_{t_0}^t dz \, e^{-(t-z)/\tau} \zeta(z) \; .
\end{align}
After averaging $\varphi(t)$ over the delta-correlated noise $\zeta(t)$ we arrive to
\begin{align}
    \langle \eta(t) \rangle &= 0 \\ 
\langle \eta(t) \eta(t^\prime) \rangle &= D \Delta(t,t^\prime) \\ 
\Delta(t,t^\prime) &\equiv \frac{1}{2 \tau} \left[ e^{-|t-t^\prime|/\tau} - e^{-(t + t^\prime - 2 t_0)/\tau}\right] \; .
\end{align}
We notice that the correlation function $\Delta$ differs from $K$ in Eq. (\ref{eq:kappa}) because of the boundary term due to the initial condition at $t_0$. In the Appendix (\ref{app:stationary_noise}) we show that a stationary noise over the finite interval $t\in[t_0,t_1]$ can be obtained by taking the initial value $\varphi_0$ from a Gaussian distribution with zero mean and variance $2 D/\tau$. The two functions equal each other once we consider the initial condition taken infinitely far in the past $t_0 \to -\infty$. 
Once we introduce the inverse operator $\Delta^{-1}(t,t^\prime)$ (that is defined and computed in the Appendix (\ref{ap:ap1})), the probability distribution of the correlated noise $\eta$ reads
\begin{align} \nonumber 
    \mathcal{P}[\eta] = \exp{ \left( -\frac{1}{2 D} \int_{t_0}^{t_1} dt \, \int_{t_0}^{t_1} dt^\prime \, \eta(t) \Delta^{-1}(t,t^\prime) \eta(t^\prime) \right) } \; .
\end{align}
After integrating over $\eta(t)$, we can write the probability of the path $\phi(t)$ as
\begin{align}
    \mathcal{P}(\phi_1,t_1,\phi_0,t_0) &= \int \mathcal{D} [\hat{\phi}] \mathcal{D} [\phi] \, e^{-S[\hat{\phi},\phi]} \\ \nonumber 
S &\equiv \int_{t_0}^{t_1} dt\,dt^\prime \mathcal{L}[\hat{\phi},\phi] \\ \nonumber 
\mathcal{L} &\equiv \hat{\phi}(t) \left[ \frac{d}{dt} \phi + r \phi - \overline{\varphi}_0(t,t_0)\right] \delta(t-t^\prime) \\ \nonumber 
&- \frac{D}{2}  \hat{\phi}(t) \Delta(t,t^\prime) \hat{\phi} (t^\prime) \; .
\end{align}
Again, since the functional integral is Gaussian, the saddle-point approximation provides the exact solution (from now on we consider as initial condition on the noise $\varphi_0\!=\!0$). The saddle-point equations are
\begin{align}
    \frac{\delta S}{\delta \phi(t)} &= - \frac{d\hat{\phi}}{dt}  + r \hat{\phi} = 0 \\
\frac{\delta S}{\delta \hat{\phi}(t)} &= \frac{d\phi}{dt}  + r\phi - D \int_{t_0}^{t_1} dt^\prime \Delta(t,t^\prime) \hat{\phi}(t^\prime) = 0 \; .
\end{align}
Once we plug the solution $\hat{\phi}(t)$ into the equation for $\phi$ we obtain 
\begin{align} \label{eq:solution_0}
    \phi(t) &= \phi_0 e^{-r(t-t_0)} + \\ \nonumber 
    & D \hat{\phi}_0 \int_{t_0}^t dt^\prime e^{-r(t-t^\prime)} \int_{t_0}^{t_1} dt^{\prime\prime} \Delta(t^\prime,t^{\prime\prime})\, e^{r (t^{\prime \prime} - t_0)} \; ,
\end{align}
where the initial value of the response field $\hat{\phi}_0$ is fixed by the condition on the endpoint $\phi(t_1)=\phi_1$. Once we set the final condition we can write
\begin{align}\label{eq:solution}
    \phi(t) &= \overline{\phi}_0(t) + \frac{\phi_1 - \overline{\phi}_0(t)}{I(t_1,t_1,t_0)} I(t,t_1,t_0)\\ \nonumber 
    \overline{\phi}_0(t) &\equiv \phi_0 \, e^{-r(t-t_0)}\\ \nonumber
    I(t,t_1,t_0) &\equiv \int_{t_0}^t dt^\prime e^{-r(t-t^\prime)} \int_{t_0}^{t_1} dt^{\prime\prime} \Delta(t^\prime,t^{\prime\prime})\, e^{r (t^{\prime \prime} - t_0)}
\end{align}
the computation of $I(t,t_1,t_0)$ is provided in Appendix (\ref{ap:compI}).
Also in this case, the analytical prediction reproduces the numerical data, as it is shown in Fig. (\ref{fig:fig2}b,\ref{fig:fig2}c) 
where we report  the comparison between the analytical solution Eq. (\ref{eq:solution}) and numerical simulations (with $\phi_0\!=\!0$, $\phi_1\!=\!1$, $D\!=\!1$, $t_0\!=\!0$, and $t_1\!=\!10$). For comparison, we have also shown the most probable path in the case of Brownian dynamics Eq. (\ref{eq:br}). 

We observe that, for small persistence time $\tau$, the trajectories tend to the Brownian path, i.e., the particle spends most of its time around the initial condition until fluctuations bring it to the final point. As $\tau$ increases, the trajectories become ballistic-like, i.e., for large $\tau$ the particle tends to follow an almost straight line connecting the initial with the final point. 

Using the numerical data, we can compute the number of paths $N(\phi_1,t_1,\phi_0,t_0)$ between $\phi_0$ and $\phi_1$, as shown by the color map in Fig. (\ref{fig:fig2}c).
The figure provides a clear visualization of the matching between the analytical solution, which reproduces the average path, and the stochastic dynamics whose trajectories accumulate around the analytical solution. 

We now compute explicitly the dynamical action along at the saddle-point. Using the saddle-point equations, we have
\begin{align} \label{eq:sp_action}
    S_{SP} = \frac{D}{2} \int_{t_0}^{t_1} dt dt^\prime \, \hat{\phi}(t) \Delta(t,t^\prime) \hat{\phi}(t^\prime)
\end{align}
where this expression differs from Eq. (\ref{eq:s_sp}) because of the boundary terms present in $\Delta(t,t^\prime)$. From the equation 
for $\hat{\phi}(t)$ we have
\begin{align}
    \hat{\phi}(t) = \hat{\phi}_0 e^{r(t-t_0)}
\end{align}
so that the action reads
\begin{align}
    S_{SP} = \frac{D \hat{\phi}_0^2}{2} \int_{t_0}^{t_1}dt dt^\prime \, e^{r(t-t_0)} \Delta(t,t^\prime) e^{r(t^\prime - t_0)} \; .
\end{align}
Again, the initial value of the auxiliary field $\hat{\phi}_0$ is fixed
by the final condition on $\phi(t)$, i.e, $\phi(t_1) = \phi_1$ meaning that 
\begin{align}
    \hat{\phi}_0 = \frac{\phi_1 - \overline{\phi}_0(t_1)}{D\, I(t_1,t_0,t_1)}
\end{align}
once we plug the expression of $\hat{\phi}_0$ into $S_{SP}$ we obtain (details about the computation are provided in Appendix (\ref{ap:compI}))
\begin{align}
    S_{SP} &= \frac{1}{2 D} \delta \phi(t_1,t_0)^2 F(t_1,t_0) \\ \nonumber 
    \delta \phi(t_1,t_0) &\equiv \phi_1 - \overline{\phi}_0(t_1) \\ \nonumber 
    F(t_1,t_0) &\equiv \frac{2 \tau r (r-\frac{1}{\tau})^2(r + \frac{1}{\tau}) e^{2t_1(r + \frac{1}{\tau})}}{B(t_1,t_0)} \\ \nonumber 
    B(t_1,t_0) &\equiv (r-\frac{1}{\tau})^2 e^{2 t_1 (r + \frac{1}{\tau})} + \frac{4 r}{\tau} e^{(t_0 + t_1)(r + \frac{1}{\tau})} \\ \nonumber
    &- r(r + \frac{1}{\tau}) e^{2(r t_1 + \frac{t_0}{\tau})} - \frac{1}{\tau} (r + \frac{1}{\tau}) e^{2(r t_0 + \frac{t_1}{\tau})} \; .
\end{align}
We now compute the stationary distribution $\mathcal{P}(\phi)$ that is obtained by performing the limit  
\begin{align}
\mathcal{P}(\phi) &\equiv \lim \limits_{\substack{t_0 \to -\infty \\ t_1 \to +\infty}} \mathcal{P}^{SP}(\phi_1,t_1,\phi_0,t_0) \\ \nonumber
\mathcal{P}^{SP} &\equiv e^{-S_{SP}}
\end{align}
where we set $\phi \equiv \lim_{t_1 \to +\infty} \phi(t_1)$. 
We thus recover the well known expression of the stationary distribution of AOUp in harmonic trap that is Gaussian with an effective spring constant depending on the correlation time of the noise \cite{Hanggi95,Maggi14,paoluzzi2022motility}
\begin{align} \label{eq:stationary}
    \mathcal{P}(\phi) &= \mathcal{N} e^{-\tilde{r}(\tau) \phi^2 / D} \\ \nonumber 
    \tilde{r}(\tau) &\equiv r(1 + r \tau ) \; .
\end{align}
Where $\mathcal{N}$ is fixed thorugh the normalization condition $\int_{-\infty}^{+\infty} d\phi \mathcal{P}(\phi) = 1$.

\subsection{Comparison with effective equilibrium dynamics}
We now compare the analytical results with those obtained considering an effective equilibrium picture. We will work within the so-called Unified Colored Noise 
(UCN) approximation \cite{Hanggi95}. For doing that, we first perform the time derivative of Eq. (\ref{eq:dyn1}) and, using Eq. (\ref{eq:oup}) and the fact that $\varphi = \dot{\phi} - f(\phi)$ we arrive to
\begin{align} \label{eq:secon_der_eq}
    \ddot{\phi} + \left[\tau^{-1} - f^{\prime}(\phi)\right]\dot{\phi} - \tau^{-1}f(\phi) = \zeta 
\end{align}
Using the path integral representation of the stochastic dynamics specified by Eq. (\ref{eq:secon_der_eq}), we can  write the path probability in the following way
\begin{align}
    \mathcal{P}(\phi_1,t_1.\phi_0,t_0) &= \int_{\phi_0}^{\phi_1} \mathcal{D}[\phi] \mathcal{D}[\hat{\phi}] \, e^{-S[\phi,\hat{\phi}]} \\ \nonumber
S &\equiv\int_{t_0}^{t_1} dt \, \mathcal{L}[\phi,\hat{\phi}] \\ \nonumber
\mathcal{L}&\equiv \frac{D}{2 \tau^2} \hat{\phi}^2 - \hat{\phi} \left[ \ddot{\phi} +\overline{\gamma}(\phi) \dot{\phi} - \tau^{-1} f(\phi) \right] \\ \nonumber 
\overline{\gamma} &\equiv \frac{1}{\tau} + V^{\prime\prime}
\end{align}
where we are considering an arbitrary force field $V(\phi)$ so that $f(\phi)=-V^\prime$. UCN assumes that $\overline{\gamma}$ is large so that we can neglect the second order time derivative. 
In order to make clear how the approximation works, we start with performing the rescaling $t = \sqrt{\tau} s$ and $\hat{\varphi} = \hat{\phi}/\sqrt{\tau}$ so that the Lagrangian density $\mathcal{L}$ now reads
\begin{align}
    \mathcal{L}[\phi,\hat{\varphi}] &= \frac{D}{2 \sqrt{\tau}} \hat{\varphi}^2 - \hat{\varphi} \left[ \ddot{\phi} + \tilde{\Gamma}(\phi) \dot{\phi} - f(\phi)\right] \\ \nonumber 
    \tilde{\Gamma} &\equiv \frac{1}{\sqrt{\tau}} + \sqrt{\tau} V^{\prime \prime} \; .
\end{align}
We now perform another change of variable by introducing $\psi = \tilde{\Gamma} \hat{\varphi}$ so that
\begin{align} \label{eq:psi}
    S[\phi,\hat{\phi}] &=\int_{t_0/\sqrt{\tau}}^{t_1 \sqrt{\tau}} ds \, \mathcal{L}[\phi,\hat{\phi}] \\ \nonumber
    \mathcal{L}[\phi,\psi] &= \frac{D}{2 \sqrt{\tau}} \left( \tilde{\Gamma}^{-1} \psi \right)^2 - \psi \left[ \tilde{\Gamma}^{-1}\ddot{\phi} + \dot{\phi} - \tilde{\Gamma}^{-1} f \right] \; .
\end{align}
In writing Eq. (\ref{eq:psi}) we have omitted the Jacobian of the transformation that is irrelevant in the harmonic case we are going to discuss. 
However, in the case of arbitrary potentials, it has to be considered.
The UCN approximation is thus obtained by performing the large effective friction limit so that $\tilde{\Gamma}^{-1}\ddot{\phi}\to0$  and we finally get
\begin{align}
    \mathcal{L}_{UCN}[\psi,\phi] &= \frac{D}{2 \sqrt{\tau}} \left( \tilde{\Gamma}^{-1} \psi \right)^2 - \psi \left[ \dot{\phi} - \tilde{\Gamma}^{-1} f \right]
\end{align}
Once we rewrite in terms of the original time scale $t=\sqrt{\tau} s$ we obtain the expression of path probability within the UCN approximation
\begin{align} \label{eq:ucn_path}
    \mathcal{P}_{UCN}(\phi_1,t_1,\phi_0,t_0) &= \int_{\phi_0}^{\phi_1} \mathcal{D}[\phi] \mathcal{D}[\psi] \,e^{-S_{UCN}[\phi,\psi]} \\ \nonumber 
S_{UCN}[\phi,\psi] &\equiv \int_{t_0}^{t_1} dt \, \mathcal{L}_{UCN}[\phi,\psi] \\ \nonumber 
\mathcal{L}_{UCN}[\phi,\psi] &= \frac{D}{2} \left( \Gamma^{-1} \psi \right)^2 - \psi \left[ \dot{\phi} + \Gamma^{-1} V^\prime \right] \\ \nonumber 
\Gamma &\equiv 1 + \tau V^{\prime \prime} \; .
\end{align}
We can now write down the self-consistency equations describing 
the instantonic trajectory for an arbitrary potential within the effective equilibrium dynamics
\begin{align}
    \frac{\delta }{\delta  \psi } \mathcal{L}_{UCN} &= D \Gamma^{-2} \psi - \dot{\phi} + \Gamma^{-1} V^\prime = 0\\
    \frac{\delta }{\delta  \phi } \mathcal{L}_{UCN}&= \frac{D}{2} \psi^2 \frac{\delta}{\delta \phi} \Gamma^{-2} + \dot{\psi} - \frac{\delta}{\delta \phi} \left[ \Gamma^{-1} V^\prime \right]= 0
\end{align}
whose solutions make sense only on region of space where the effective friction $\Gamma$ is positive. 
Focusing our attention to the harmonic potential $V\!=\!r \phi^2 /2$, one has $V^{\prime\prime}\!=\!r$ and the equations can be written as
\begin{align}
    & \frac{D}{(1 + r\tau)^{2}} \psi - \dot{\phi} - \frac{r}{1 + r\tau}\phi = 0 \\ 
    & \dot{\psi} - \frac{r}{1 + r \tau} \psi = 0 
\end{align}
as a result, the most probable path $\phi_{UCN}(t)$ is the same as the Brownian explored before that brings to Eq. (\ref{eq:br}) with the effective elastic constant $\hat{r} = r/(1+\tau r)$
\begin{align} \label{eq:ucn_traj}
    \phi_{UCN}(t) &= \overline{\phi}_0(t) + \frac{\phi_1 - \overline{\phi}_0(t)}{\sinh \left[ \hat{r} (t_1 - t_0)\right]} \sinh \left[ \hat{r} (t - t_0)\right] \\ \nonumber   
    \overline{\phi}_0(t)&= \phi_0 e^{-\hat{r} (t - t_0)} \; .
\end{align}
As we discussed before, the corresponding stationary distribution can be obtained by performing the limit $t_{0,1} \to \mp \infty$ with the condition $\phi = \lim_{t_1 \to \infty} \phi(t_1)$ and then we have $\mathcal{P}_{UCN}(\phi) \propto e^{-\tilde{r}(\tau) \phi^2 / D}$
that is the same we obtained in Eq. (\ref{eq:stationary}) (it is well known that in the case of harmonic potential the stationary distribution obtained within UCN matches the analytical one \cite{Hanggi95}). 

In Fig. (\ref{fig:fig3}a) we show a comparison between the trajectories obtained by the analytical solution and the ones within UCN.  
As one can see, although the stationary distributions are the same, the dynamics most probable trajectories within the two dynamics are considerably different. If we look at $\phi - \phi_{UCN}$ as a function of time, as it is shown in Fig. (\ref{fig:fig3}b), the effective and the actual dynamics matches only for early times in the small $\tau$
limit, i.e., only on times $t \ll r \tau$, while they differ almost everywhere as $\tau$ increases. This is because UCN replaces the actual dynamics that is governed by two time scales, i.e., the characteristic time of the deterministic force $r^{-1}$ and the characteristic time of noise $\tau$, with only one time scale, that we call $\tau_{UCN}$, that is the sum of the two, i.e., $\tau_{UCN} = \tau + r^{-1}$.
In the Appendix (\ref{app:two_points}) we show this by performing the computation of the two-point correlation functions in the actual and UCN dynamics. From their computation, it turns out that UCN dynamics can reproduce the exact dynamics only for short times and only in the limit $\tau \to 0$.

\section{Discussion}
We have computed the most probable trajectory followed by an AOUp immersed in harmonic traps. The analytical computation is made possible because the functional integration involves Gaussian integral so that the saddle point approximation becomes exact.

We first performed the computation in the extended space that counts as degrees of freedom particle's position $\phi$ and the self-propulsion $\varphi$, i.e., the extended Markovian dynamics. In this case, we can compute the trajectory with arbitrary initial and final conditions on position and self-propulsion velocity. We stress that the knowledge of analytical expression for the most probable path that takes into account also self-propulsion might be useful for gaining insight into the self-propulsion mechanisms in experiments.

Then, we computed the trajectory of the non-Markovian dynamics with arbitrary conditions on particle's position. We have tested both results against numerical simulations, obtaining perfect agreement between numerics and theory. As a general result, the dynamical action at the saddle point is quadratic in the response field $\hat{\phi}$. This fact provides an intuitive physical interpretation of the response field as an external field that fixes the boundary condition on the endpoint. Finally, we obtained the stationary distribution by performing the large-time limit of the probability distribution at the saddle-point.

We have also computed the trajectories followed by the effective equilibrium dynamics within the UCN approximation scheme. We obtained that effective equilibrium trajectories differ substantially almost everywhere from those of the actual dynamics. We observe only a partial agreement between the two dynamics in the small $\tau$ limit on time scale $t \ll \tau$ (at fixed $r$).

In the present paper, we focused our attention on the harmonic case where we can compute analytically the path integral. However, we considered a general formalism that is suitable for the computation of istantonic trajectories in presence of arbitrary external potentials. In particular, Eqs. (\ref{eq:saddle_point}) have been obtained in the small noise limit
so that their solution with arbitrary initial conditions provides uphill and downhill trajectories in the regime where barrier jumping is possible only because of persistent fluctuations. Part of this program has already been followed and it brought to the computation of the Onsager-Machlup action (in the small and large $\tau$ limit), as well as the prefactor of the path integral for estimating the escape rate from the local minima of a double-well potential \cite{PhysRevA.41.657,PhysRevA.42.1982}. However, the shape of the typical trajectories and how it changes for different values of self-propulsive forces at the endpoint of the trajectory remain unexplored.
Moreover, while in the computation of the transition rate between minima one consider the limit $t_0\to-\infty$ and $t_1\to\infty$ motivated by the fact one is interested in rare jump events between two stationary point, i.e., $\phi(-\infty)=\phi_0$ and $\phi(\infty)=\phi_1$, the formalism developed here consider arbitrary initial and final conditions.

We also stress that the theoretical set-up we employed for addressing the computation of the trajectories of the extended dynamics, i.e., $(\phi_0,\varphi_0) \to (\phi_t,\varphi_t)$ can be generalized to different situations. To be more specific, Eq. (\ref{eq:linear_sys})
holds for a vast class of dynamical systems that count many degrees of freedom coupled linearly with each other but, in principle, in a non-reciprocal way, this is because the computation holds for generic to non-symmetric $\boldsymbol{M}$. For instance, one can consider an arbitrary Gaussian correlated noise $K(|t|/\tau)$ whose Fourier Transform can be written in power series of $\tau^m$, with $m>1$, (see Ref. \cite{PhysRevA.41.644} for details), and then, by embedding it into an opportune cascade of $m$ Ornstein-Uhlenbeck processes, so that $\boldsymbol{M}$ is a $(m+1) \times (m+1)$ non-symmetric matrix. 
Finally, in writing Eq. (\ref{eq:noise_matrix}) we did not consider the effect of a thermal noise that can be taken into account through an opportune choice of $\boldsymbol{D}$. It might be also interesting to explore the computation of the most probable path within other theoretical frameworks based on path integrals as in the case of the Doi-Peliti formalism that has been recently considered for addressing Active Matter dynamics \cite{garcia2021run,PhysRevE.103.062105,pruessner2022field}.

\subsection*{Acknowledgments}
M.P. has received funding from the European Union's Horizon 2020 research and innovation program under the MSCA grant agreement No 801370
and by the Secretary of Universities 
and Research of the Government of Catalonia through Beatriu de Pin\'os Program Grant No. BP 00088 (2018).

\appendix 
\section{Computation of $\Delta^{-1}(t,t^\prime)$} \label{ap:ap1}
In this section, we compute the inverse of the operator $\Delta(t,t^\prime)$ defined in the main text. 
We can define the inverse operator $\Delta^{-1}(t,t^\prime)$ in the following way
\begin{align} \label{eq:invDelta}
    \int_{t_0}^{t_1} dt^{\prime \prime} \Delta^{-1}(t,t^{\prime\prime}) \Delta(t^{\prime\prime},t^\prime) = \delta(t-t^\prime) \; ,
\end{align}
so that we can define the probability distribution of the correlated noise $\eta$ as follows (we perform the replacement $\tau=\gamma^{-1}$)
\begin{align}
    \mathcal{P}[\eta] = \exp{ \left( -\frac{1}{2 D} \int_{t_0}^{t_1} dt \, \int_{t_0}^{t_1} dt^\prime \, \eta(t) \Delta^{-1}(t,t^\prime) \eta(t^\prime) \right) } \; .
\end{align}
In order to compute $\Delta^{-1}$  we rewrite $\Delta$ as follows
\begin{align}
    \Delta(t,t^\prime) &= \frac{\gamma}{2} \left[ e^{-\gamma (t - t^\prime)} \theta(t-t^\prime) \right. \\ \nonumber 
    &+ \left.e^{\gamma(t-t^\prime)} \theta(t^\prime - t) - e^{-\gamma(t+t^\prime - 2 t_0)}\right] \end{align}
By performing the time derivatives of $\Delta$ we get
\begin{align}
    \frac{d}{dt}     \Delta(t,t^\prime) &= \frac{1}{2} \left[ -e^{-\gamma(t-t^\prime)}\theta(t-t^\prime) \right. \\ \nonumber
    &\left. +  e^{\gamma(t-t^\prime)} \theta(t^\prime - t) + e^{-\gamma(t + t^\prime -2 t_0)} \right] \\ \nonumber 
    \frac{d^2}{dt^2} \Delta(t,t^\prime) &= \frac{\gamma}{2} \left[ e^{-\gamma(t-t^\prime)}\theta(t-t^\prime) \right. \\ \nonumber 
    &\left. + e^{\gamma(t-t^\prime)} \theta(t^\prime - t) - e^{-\gamma(t + t^\prime + 2t_0)} \right] - \delta(t-t^\prime) \\ \nonumber 
    &= \gamma^2 \Delta(t,t^\prime) - \delta(t-t^\prime) \; ,
\end{align}
and thus we obtain
and thus we arrive to the equation
\begin{align}
    \left[ -\frac{d^2}{dt^2} + \gamma^2 \right] \Delta(t,t^\prime) &= \delta(t-t^\prime) \\ \nonumber 
    \Delta(t_0,t^\prime) &= 0
\end{align}
because of Eq. (\ref{eq:invDelta}), one has
\begin{align}
    \Delta^{-1}(t,t^\prime) &= \left[ -\frac{d^2}{dt^2} + \gamma^2 \right] \delta(t - t^\prime) \; .
\end{align}

\section{Computation of $I(t,t_1,t_0)$ and $H(t_1,t_0)$} \label{ap:compI}
We have to evaluate the following integral
\begin{align} \label{eq:compI}
    I(t,t_0,t_1) &\equiv  \int_{t_0}^t dw \, J(w,t_0,t_1) e^{-r(t-w)}
\end{align}
where we have defined the quantity $J(w,t_0,t_1)$ as follows
\begin{align}
    J(w,t_0,t_1)&\equiv \int_{t_0}^{t_1} dw^\prime \, \Delta(w,w^\prime) \, e^{r(w^\prime - t_0)}
\end{align}
once we perform the replacement $\gamma = \tau^{-1}$, we obtain
\begin{align} \label{eq:compJ}
    J(w,t_0,t_1)&=\frac{\gamma}{2} e^{r (w-t_0)} \left[-\frac{e^ {-(\gamma +r) (w-t_0)}}{\gamma +r} \right. \\ \nonumber 
    &\left. +\frac{e^{ (r-\gamma ) (t_1-w) }}{r-\gamma }-\frac{2 \gamma }{(\gamma +r) (r-\gamma )}\right] \\ \nonumber
    &-\frac{\gamma  e^{-\gamma  (w-t_0) } }{2 (r-\gamma )} \left[ e^{ (r-\gamma ) (t_1-t_0) } -1 \right]
\end{align}
once we plug Eq. (\ref{eq:compJ}) into Eq. (\ref{eq:compI}) we finally obtain
\begin{align}
       I(t,t_0,t_1) &= -\frac{\gamma}{2 r (r-\gamma )^2 (\gamma +r)}  e^{-r (t+ t_0)-\gamma  (t+ t_1)} \times
        \\ \nonumber 
&\left[ (\gamma + r) \left( r e^{r (t+ t_1)+2 \gamma  t_0} + \gamma e^{2 r t_0+\gamma  (t+t_1)} \right) \right. \\ \nonumber 
       & \left.  -2 \gamma  r \left( e^{r ( t_0+ t_1)+\gamma  (t+ t_0)} + e^{r (t+ t_0)+\gamma  ( t_0+ t_1)} \right) \right. \\ \nonumber 
       &\left.+
       (r-\gamma ) \left( \gamma e^{2 r t+\gamma  (t+ t_1)} - r e^{r (t+ t_1)+2 \gamma  t}\right) 
       \right] \; .
\end{align}
We now proceed with the computation of $F(t_1,t_2)$ defined as follows
\begin{align} \label{eq:f}
    F(t_1,t_0) &\equiv \frac{H(t_1,t_0)}{I(t_1,t_0,t_1)^2} \\ 
    H(t_1,t_0) &\equiv \int_{t_0}^{t_1} dw dw^\prime \, e^{r(w-t_0) } \Delta(w,w^\prime) e^{r (w^\prime - t_0)} \; ,
\end{align}
The computation of $H(t_1,t_0)$ can be done using $J$ so that
\begin{align}
    H(t_1,t_0) = \int_{t_0}^{t_1} dw \, e^{r(w-t_0) } J(w,t_0,t_1)
\end{align}
and we finally obtain
\begin{align} \label{eq:h}
    H(t_1,t_0) &= \frac{\gamma}{2 r (r-\gamma)^2 (r+\gamma)} \times \\ \nonumber 
    &\left[ (r-\gamma)^2 e^{2r (t_1-t_0)} + 4 r \gamma e^{-(t_0-t_1)(r-\gamma)} \right. \\ \nonumber  &\left.-r(r+\gamma)e^{-2(t_0-t_1)(r-\gamma) - \gamma (r+\gamma)}\right] \; .
\end{align}
Once we plug Eq. (\ref{eq:h}) into Eq. (\ref{eq:f}) we get
\begin{align}
    F(t_1,t_0) &= \frac{2 r}{\gamma} \frac{(r-\gamma)^2(r + \gamma) e^{2 t_1 (r + \gamma)}}{B(t_1,t_0)} \\ 
    B(t_1,t_0) &\equiv (r - \gamma)^2 e^{2t_1(r + \gamma)} + 4 r \gamma e^{(t_0 + t_1)(r + \gamma)} \\ \nonumber 
    &-r(r + \gamma) e^{2(r t_1 + \gamma t_0)} - \gamma (r + \gamma) e^{2(r t_0 + \gamma t_1)} \; .
\end{align}
\section{Numerical simulations} \label{ap:ap2}
\begin{figure}[!th]
\centering\includegraphics[width=.4\textwidth]{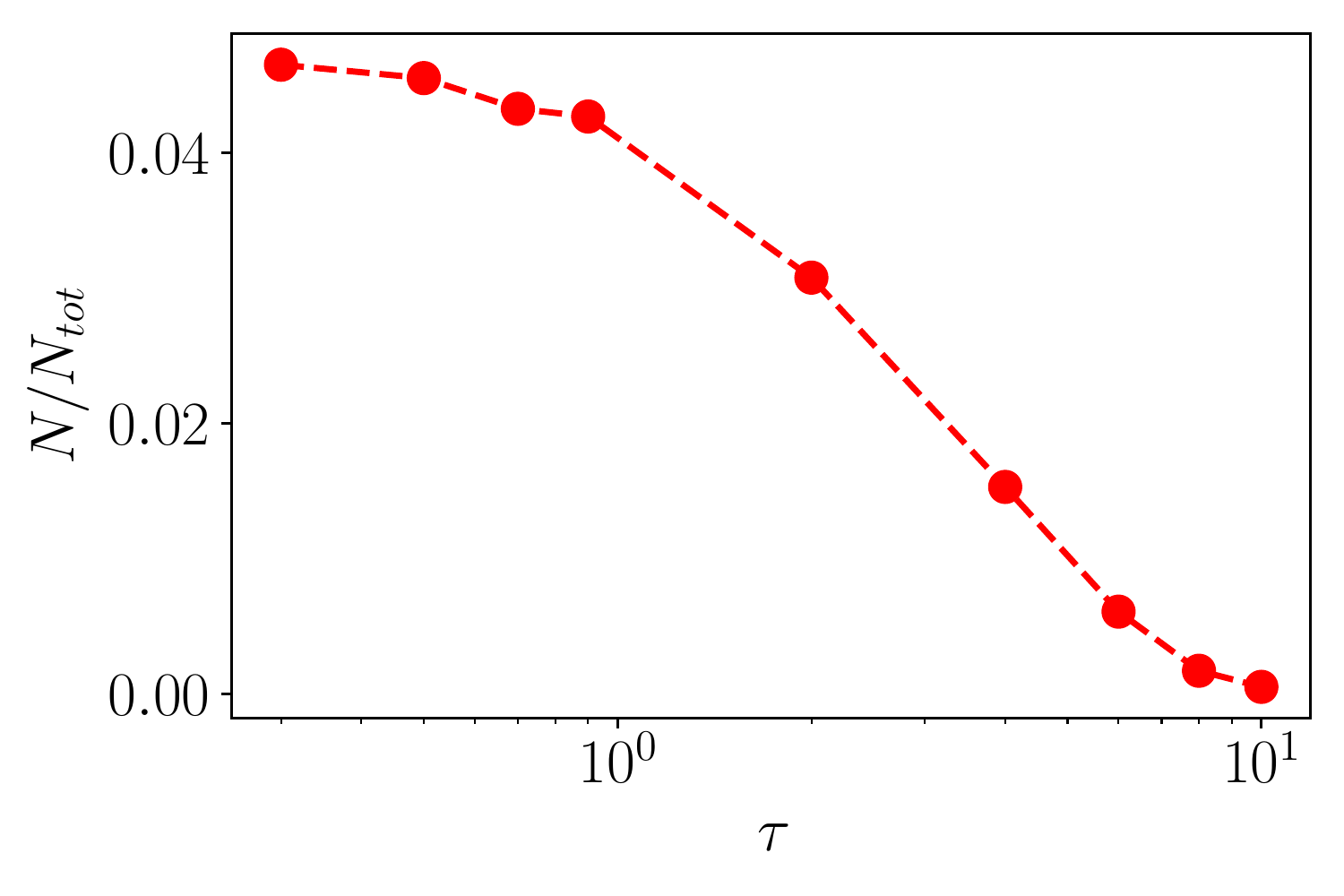}
\caption{Fraction of trajectories around the final point $\phi_1\!=\!\phi(t_1)$, with $\phi_1\!=\!1$, $t_1\!=\!10$, and $\phi_0\!=\!\phi(0)\!=\!0$, as a function of $\tau$.}
\label{fig:fig_app}
\end{figure}
The analytical prediction have been tested against numerical data produced by solving numerically (using Euler scheme with integration time-step $\Delta t=10^{-3}$) the stochastic dynamics give by
\begin{align} \label{eq:dyn_sim}
    \dot{\phi}    &= -r \phi + \varphi \\ 
    \dot{\varphi} &= -\tau^{-1} \varphi + \zeta  
\end{align}
with the initial conditions $\phi_0=\phi(t_0)$ and $\varphi_0=\varphi(t_0)$. We have considered $D=1$, $\tau\in[0.3,10]$, and $r=1$. $\zeta$ represents the usual Gaussian noise with zero mean and variance $2D/\tau^2$. The solid lines in Fig. (\ref{fig:fig2}a) and Fig. (\ref{fig:fig2}b) have been obtained by considering $N_{tot}=10^5$ independent runs. The initial conditions at $t_0=0$ are $\varphi_0=\phi_0=0$. As final time, we set $t_1=10$ where we look for trajectories such as $\phi(t_1)=\phi_1=1$.
We have thus selected the $N$ trajectories satisfying the final conditions $\phi(t_1)=t_1$ (and $\varphi(t_1)=\varphi_1$ for producing Fig. (\ref{fig:fig2}a)) and averaged over them for obtaining the average path connecting the initial with the final point. 
In Fig. (\ref{fig:fig_app}) we show the dependency on the correlation time $\tau$ of the fraction $N/N_{tot}$ of trajectories that satisfy the final condition $\phi_1$. 

\section{Computation of the two-point function} \label{app:two_points}
In this section we compute the two point-function and we compare the exact dynamics with those obtained within UCN.
We start fro the equations of motion for the two degrees of freedom
\begin{align} \label{app:eq_dyn_again}
    \dot{\phi} &= -r \phi + \varphi \\ \nonumber 
    \dot{\varphi} &= -\tau^{-1} \varphi + \zeta \\ \nonumber 
    \langle \zeta(t) \rangle &= 0 \, , \;\; \langle \zeta(t) \zeta(t^\prime) \rangle = 2 \tau^{-2} D \delta(t - t^\prime) 
\end{align}
we consider a stationary correlated noise $\varphi$ (as discussed in the next section), a condition that is satisfied once we take the initial condition $\varphi_0=\varphi(t_0)$ from a Gaussian distribution 
\begin{align}
    \langle \varphi_0 \rangle &= 0 \, , \, \langle \varphi_0^2 \rangle = 2 \tau^{-1} D \; .
\end{align}
Once we perform the time derivative of the first equation in Eq. (\ref{app:eq_dyn_again}), 
we arrive at the second-order stochastic differential equation
\begin{align}
    \ddot{\phi} &+ \gamma \dot{\phi} + \kappa \phi = \zeta \\ \nonumber 
    \gamma &\equiv r + \tau^{-1} \, , \; \kappa \equiv r \tau^{-1} 
    \\ \nonumber 
    \phi(t_0) &= \phi_0 \\ \nonumber 
    \left. \dot{\phi} \right|_{t=t_0} &= -r \phi_0 + \varphi_0 \; .
\end{align}
We now introduce the green function $G(t,t^\prime)$ given by the solution of the equation
\begin{align}
    \left[ \partial_t^2 + \gamma \partial_t + \kappa \right] G(t,t^\prime) = \delta(t - t^\prime) 
\end{align}
with solution
\begin{align}
    G(t,t^\prime) &= \frac{e^{-\lambda_+ (t - t^\prime)} - e^{-\lambda_- (t - t^\prime )}}{\lambda_- - \lambda_+} \theta(t - t^\prime) \\ \nonumber 
    \lambda_{\pm} &= \frac{\gamma \pm \sqrt{\gamma^2 - 4 \kappa} }{2}
\end{align}
and, once we consider the initial conditions $\phi_0$ and $\dot{\phi}_0$, the solution for $\phi$ reads
\begin{align}
    \phi(t) &= a \, e^{-\lambda_+ t} + b \, e^{-\lambda_- t} + \int_{t_0}^t ds \, G(t,s) \zeta(s) \\ \nonumber 
    a &= \frac{\lambda_- \phi_0 + \dot{\phi}_0}{\lambda_- - \lambda_+} \\ \nonumber 
    b &= \frac{\lambda_+ \phi_0 + \dot{\phi}_0}{\lambda_- - \lambda_+} \; .
\end{align}
The presence of two time scales, one given by the correlation time of the noise $\tau$ (the noise time-scale), and the second one the characteristic time of the harmonic potential $r^{-1}$ (the deterministic time-scale), is reflected by $\lambda_{\pm}$. For $r<1/tau$, one has $\lambda_+=\tau^{-1}$ and $\lambda_-=r$. In the other case, i.e., $r>\tau^{-1}$, one has $\lambda_+=r$ and $\lambda_-=\tau^{-1}$.
The two-point function (with $t \geq t^\prime$) is
\begin{align}
    \langle \phi(t) \phi(t^\prime) \rangle &= \frac{D}{2 \tau^2} \frac{1}{\lambda_+^2 -\lambda_-^2} \left[ \lambda_{-}^{-1} e^{-\lambda_- (t - t^\prime)} - \lambda_{+}^{-1} e^{-\lambda_+ (t - t^\prime)} \right] \\ \nonumber 
&+ \frac{1}{(\lambda_+ - \lambda_-)^2}g(t,t^\prime,t_0) \\ \nonumber 
g(t,t^\prime,t_0) &\equiv A_0^+ e^{-\lambda_+ (t + t^\prime - 2 t_0)} + A_0^- e^{-\lambda_- (t + t^\prime - 2 t_0)}  \\ \nonumber 
&+A_1 \left[ e^{-\lambda_-(t-t_0) - \lambda_+ (t^\prime - t_0)} + e^{-\lambda_-(t^\prime-t_0) - \lambda_+ (t - t_0)}\right] \\ \nonumber 
A_0^{\pm} &\equiv (\lambda_\pm - r)^2 \phi_0^2  + \frac{D}{2 \tau}(1 - \frac{1}{\tau \lambda_\mp}) \\ \nonumber 
A_1 &\equiv (\lambda_+ -r)(\lambda_- - r) + \frac{ D}{2\tau} (1 - \frac{2}{\tau (\lambda_+ + \lambda_-)})\; .
\end{align}
Once we plug the expressions of $\lambda_{\pm}$ we obtain
\begin{align}
   \langle \phi(t) \phi(t^\prime) \rangle &= \frac{D}{2 r} \frac{1}{1 - \tau^2 r^2} \left[ e^{-r|t-t^\prime|} - \tau r e^{-\frac{1}{\tau}|t-t^\prime|}\right] \\ \nonumber 
   &+ g(t,t^\prime,t_0) \\ \nonumber 
   g(t,t^\prime,t_0) &\equiv \frac{D}{2} \frac{\tau}{1 - \tau^2 r^2} A_0 \\ \nonumber 
   &+ \left[ \phi_0^2 - \frac{D}{2 r} \frac{1}{1 - \tau r}\right] e^{-r(t+t^\prime -2 t_0)} \\ \nonumber 
   A_0 &\equiv e^{-r(t - t_0) - \frac{1}{\tau}(t^\prime - t_0)} + e^{-r(t^\prime - t_0) - \frac{1}{\tau}(t - t_0)} 
\end{align}
In contrast, using the rescaled time $s=t / \sqrt{\tau}$, within UCN the dynamics is governed by
\begin{align}
    &\gamma \partial_s \phi + r \phi = \zeta \\ \nonumber 
    \gamma &= \sqrt{\tau} r  + \frac{1}{\sqrt{\tau}} \\ \nonumber 
    &\langle \zeta(s) \zeta(s^\prime) \rangle = \frac{2 D}{\sqrt{\tau}} \delta(s-s^\prime)
\end{align}
Once we go back to time variable $t$,
the two-point function is given by
\begin{align}
    \langle \phi(t) \phi(t^\prime) \rangle_{UCN} &= \frac{D}{2 r} \frac{1}{\tau r +1 }  e^{-\frac{r}{1 + \tau r} |t - t^\prime|} \\ \nonumber 
    &+    \left( \phi_0^2 - \frac{D}{2 r} \frac{1}{1 + \tau r}\right) e^{-\frac{r}{r \tau + 1} (t + t^\prime - 2 t_0)} \; .
\end{align}
The relaxation dynamics within UCN evolves on a single time-scale that it the sum of the two time scales 
\begin{align}
    \frac{r\tau + 1}{r} = \tau + \frac{1}{r} \; .
\end{align}
As a consequence, the two dynamics are equivalent only in the limit $\tau r \ll 1$. Meaning that, if we fix $r$ to a constant, the two dynamics agree with each others only in the $\tau \to 0$ limit.
Looking at the opposite limit, i.e., $\tau \to \infty$ for fixed $r$, the agreement between exact and approximated dynamics hold only for times $t,t^\prime \gg 1/r$. 

In the limit $t_0 \to -\infty$ the correlation functions are time translational invariant since they depend only on the difference $|t - t^\prime|$. In this limit, when we consider $\tau \to 0$, we have
\begin{align}
    \langle \phi(t) \phi(t^\prime)\rangle &\sim \frac{D}{2 r} e^{-r |t-t^\prime|} \left[ 1 + \tau^2 r^2 + O(\tau^4)\right] \\ \nonumber 
        \langle \phi(t) \phi(t^\prime)\rangle_{UCN} &\sim \frac{D}{2 r} e^{-r |t-t^\prime|} \left[ 1 - \tau r  + \tau r^2 |t - t^\prime| + O(\tau^4)\right] 
\end{align}
meaning that UCN provides only an asymptotic $O(1)$ approximation to the exact solution as $\tau \to 0$.
\section{Stationary Noise} \label{app:stationary_noise}
We start from the expression of $\varphi(t)$ that is given in Eq. (\ref{eq:sol_varphi}) that is
\begin{align}
    \varphi(t) &= e^{-\gamma (t-t_0)} \varphi_0 + \eta(t)
\end{align}
and thus we have
\begin{align} 
    \langle \varphi(t) \rangle &= e^{-\gamma (t - t_0)} \langle \varphi_0 \rangle \\ \nonumber
    \langle \varphi(t) \varphi(t^\prime) \rangle &= 2 D \gamma^2 e^{-\gamma |t - t^\prime| } + \left[ \langle \varphi_0^2 \rangle - 2 D \gamma^2\right] e^{-\gamma (t + t^\prime - 2 t_0)} \; .
\end{align}
Once we require that $\varphi(t)$ has to be a stationary process,
we have to options: The first one is to take the initial condition at an initial time infinite in the past, i.e., $t_0 \to -\infty$, in alternative, if we want to keep $t_0$ finite, we can extract $\phi_0$ from a Gaussian distribution with
\begin{align}
    \langle \varphi_0 \rangle &= 0 \\ 
        \langle \varphi_0^2 \rangle &= 2 D \gamma^2 \; .
\end{align}

\bibliography{mpbib}
\bibliographystyle{rsc}

\end{document}